\documentclass[aps,floats,floatfix,amssymb,amsmath,prb,nofootinbib,twocolumn,superscriptaddress]{revtex4-2}

\usepackage{physics} 
\usepackage[T1]{fontenc}
\usepackage[latin1]{inputenc}
\usepackage{graphicx}
\usepackage{hyperref}
\usepackage{babel}
\usepackage{amsfonts}
\usepackage{color}
\usepackage{capt-of}
\usepackage[font=small]{caption}

\usepackage{multirow}
\usepackage{xspace}
\usepackage[normalem]{ulem}

\usepackage{ragged2e} % for \justifying

\newcommand{\red}{\color{red}}

\begin{document}
\title{Precursors to Anderson localization in the Holstein model: Quantum and quantum-classical solutions}

\author{P. Mitri\'c}
\affiliation{Institute of Physics Belgrade, University of Belgrade, 
Pregrevica 118, 11080 Belgrade, Serbia}

\author{V. Dobrosavljevi\'c}
\affiliation{Department of Physics and National High Magnetic Field Laboratory, Florida State University, Tallahassee, Florida, USA}

\author{D. Tanaskovi\'c}
\affiliation{Institute of Physics Belgrade, University of Belgrade, 
Pregrevica 118, 11080 Belgrade, Serbia}

\begin{abstract}

We calculate the frequency-dependent mobility of the Holstein polaron in one dimension near adiabatic limit using the method based on dynamical quantum tipicality, as well as the quantum-classical method. The agreement between fully quantum and quantum-classical solutions is very good. The most prominent feature is the appearance of a zero-frequency peak in the mobility, in addition to the displaced peak associated to the precursors of Anderson localization. The zero-frequency peak cannot be obtained within the phenomenological transient localization approach, which is often used in a semiquantitative description of charge transport in quasi-one-dimensional organic semiconductors.

\end{abstract}

\maketitle

%%%%%%%%%%%%%%%%%%%%%%%% INTRODUCTION %%%%%%%%%%%%%%%%%%%%%%%%%%%%%%%%%%%%%%
{\it Introduction.}
Charge mobility is a key quantity which determines the optoelectronic properties of molecular organic semiconductors \cite{Coropceanu2007,Zhugayevych2015,Fratini2020}.
Weak van der Waals forces between the organic molecules lead to strong lattice thermal fluctuations and the charge transport in pure samples near the room temperature is dominated by the electron-phonon scattering.  It is common to distinguish between the local electron-phonon interaction and its nonlocal part, which are often modeled by the Holstein and Peierls Hamiltonian, respectively. However, in many cases the electron-phonon scattering is too strong to be treated by perturbative methods \cite{Fratini2016}, the charge transport is in between the band and the hopping limit \cite{Troisi_review2011}, and a reliable quantum calculation of the dc mobility is lacking, even within these two simplified models. The real-frequency calculations are often restricted to lattices that are not sufficiently large \cite{Schubert_2005,Fratini_PRL2024}, while the effectiveness of the imaginary-axis quantum Monte Carlo (QMC) calculations \cite{Mishchenko_2015} is limited by the ill-defined analytical continuation \cite{Meyer_2011}, since a small difference in the imaginary-time current-current correlation function can correspond to a substantial difference in conductivity.
\cite{Vucicevic_2019,Jankovic_2024,Buividovich_2024}.
%{\it {comment: finite-size effects and analytical continuation; the bandlike limit and hopping limit; difficult to determine how reliable are quantum solutions}}

An important insight into the charge transport in quasi-one-dimensional organic semiconductors is obtained by the phenomenological transient localization (TL) theory \cite{Fratini2016,Ciuchi_PRB2011}. It starts from the observation that on short time scales, much smaller than the period of lattice oscillation $2\pi/\omega_0$, the ions displacements can be considered as {\it almost static} and randomly distributed (with appropriate probability distribution). Hence, on these time scales,  the charge transport can be described by physics of an electron moving thorough the statically disordered environment. In a model of {\it fully static} disorder, i.e. in the Anderson model (AM), the electron wave functions would be localized in low dimensions. However, the phonons are a source of dynamical disorder, causing the inelastic scattering which breaks the localization at longer time scales. In TL approach, which is not restricted to a particular model, the inelastic scattering is accounted in a relaxation time approximation through the phenomenological parameter $\tau_{\mathrm{in}} \sim 1/\omega_0$ which, at long times, modifies the current-current correlation function corresponding to the AM, $C_{jj}^{\mathrm{TL}}(t)=C_{jj}^{\mathrm{AM}}(t)e^{-t/\tau_{\mathrm{in}} }$, leading to nonzero dc mobility. 
The frequency-dependent mobility $\mu(\omega)$ features a so-called displaced Drude peak (DDP) \cite{Fratini_PRL2024}, which is here a signature of precursors to the Anderson localization. Its appearance is, however, a more general phenomenon. DDP is observed in different physical systems, including some bad metals, and its origin is a subject of several recent studies \cite{Pustogow2021,Fratini_2021,Aydin_PRL2024,Aydin_2024}.

%
%The TL approach can be equally easy applied both to the Peierls and to the Holstein model.

Another popular approach to charge dynamics is given by the quantum-classical (QC) methods \cite{Tully_2023,Troisi_PRL2006,Fetherolf_PRX2020}. Here, the electron part of the Hamiltonian is treated quantum-mechanically, whereas the lattice vibrations are treated classically. The back-action of the electron to lattice vibrations is usually included through the Ehrenfest equations. In this case, the total energy of the system is conserved, but the electron energy increases with the propagation time. At large times, the electron energy is not distributed according to the Boltzmann statistics at temperature $T$, but instead follows the distribution corresponding to infinite temperature. This feature has led to the conclusion that the QC method gives a divergent diffusion constant $D(t)$ at $t \rightarrow \infty$, implying that it cannot be used to calculate the dc mobility \cite{Fratini2016,Ciuchi_PRB2011}. However, a very recent work \cite{Runeson_2024} on the one-dimensional (1D) Peierls model for the parameter models corresponding to rubrene finds that $D(t)$ features an upturn at $t \approx 1/\omega_0$, but reaches a plateau for $t\sim 2\pi/\omega_0$. The upturn in $D(t)$ corresponds to a zero-frequency peak in the frequency-dependent mobility $\mu(\omega)$ of width $\omega_0$, which is absent in the TL solution. Furthermore, it was shown that a similar result for $D(t)$ can be obtained within the newly developed mapping approach to surface hopping \cite{Runeson_2024,Runeson_2023} which conserves the electron energy, giving an important support to the Ehrenfest dynamics result.
Still, there are important questions that have remained open. In particular, how applicable is the QC approximation? Will the zero frequency peak appear also in the fully quantum solution? Does the answer depend on a specific model and the parameter regime?

To answer these questions, in this Letter we focus on the 1D Holstein model, where due to the recent methodological advances, we can find a fully quantum solution representative of the thermodynamic limit, at least in certain parameter regimes. Specifically, we will use the method based on dynamical quantum typicality (QT) \cite{Mitric_preprint} which is complemented by the publicly available solutions of the hierarchical equations of motion (HEOM) \cite{Jankovic_2023,Jankovic_2024,Jankovic_Zenodo8}. We find very good agreement with the numerically cheaper QC calculations, both featuring non-monotonous diffusion $D(t)$ with a plateau at large times. Both the fully quantum and QC solution feature a zero-frequency peak in $\mu(\omega)$, apart from the finite-frequency DDP which appears as a precursor to the Anderson localization at short time scale. Our results, when combined with those for the Peierls model \cite{Runeson_2024,Buividovich_2024}, indicate that the zero frequency peak in mobility appears both in the models with local and nonlocal electron-phonon coupling.

{\it Model and methods.} The 1D Holstein model is given by the Hamiltonian
\begin{align} \label{eq:Holstein_hamiltonian}
H =& -t_0 \sum_i \left( c_i^\dagger c_{i+1} + \mathrm{h.c.} \right) \nonumber\\
   & -g \sum_i n_i \left( a_i^\dagger + a_i \right) + \omega_0 \sum_i a_i^\dagger a_i .
\end{align}
%=& H_{\mathrm{el}} + H_{\mathrm{el-ph}} + H_{\mathrm{ph}} \nonumber\\
%
Here, $t_0$ is the hopping parameter, $c_i^\dagger$ ($a_i^\dagger$) is the electron (phonon) creation operator, $n_i = c_i^\dagger c_i$, and we assume a single electron in the band as appropriate for low doped semiconductors. The electron-phonon coupling constant is denoted by $g$, the phonon frequency by $\omega_0$, and we also introduce a convenient dimensionless quantity $\lambda = g^2 / (2\omega_0 t_0)$. We set $t_0$, $\hbar$, $k_B$, $e$, and the lattice constant to one. %, and the ion mass are set to one.}
Within the Kubo linear-response formalism \cite{Kubo_1957,Mahan,Bertini_2021}, the time-dependent diffusion constant, $D(t)  = \int_0^t dt' \mathrm{Re} C_{jj}(t')$, and the frequency-dependent mobility,
\begin{equation}
   \mu(\omega) = \frac{2 \tanh \left( \frac{\beta \omega}{2} \right)}{\omega}
  \int_0^\infty dt \cos(\omega t) \mathrm{Re} C_{jj}(t) ,
\end{equation}
are obtained from the current-current correlation function $C_{jj}(t)=\langle j(t) j(0)\rangle$, where ${j=i t_0 \sum_i \left( c_{i+1}^\dagger c_i - c_i^\dagger c_{i+1} \right)}$. The dc mobility is given by the Einstein relation ${\mu_{\mathrm{dc}} = D(\infty)/T}$.

We solve the Hamiltonian given by  Eq.~\ref{eq:Holstein_hamiltonian} by the methods which, in certain parameter regimes, give numerically exact result for $C_{jj}(t)$ representative of the thermodynamic limit. The QT method \cite{Heitmann_2020,Jin_2021} is presented in detail in Ref.~\cite{Mitric_preprint}. Its application is mostly restricted by the computer memory since the Hilbert space grows rapidly with the total number of phonons and lattice sites that we take into account. We use it for intermediate and strong coupling, where one can eliminate the finite-size effects within available memory. For the solution at high temperature and for weak coupling, we use the HEOM results from the literature \cite{Jankovic_2023,Jankovic_Zenodo8}. At low temperature we need a much longer chain, but for weak interaction we can calculate the mobility within the dynamical mean-field theory (DMFT) \cite{Ciuchi1997}. 
In Ref.~\cite{Mitric_2022} we showed that, rather surprisingly, the DMFT gives nearly exact single particle properties within the Holstein model, in arbitrary number of dimensions and the corresponding exact self-energy is nearly local. The bubble term for conductivity then almost coincide in the HEOM and DMFT solutions \cite{Jankovic_2024}. Furthermore, we showed that the vertex corrections to conductivity in the Holstein model vanish in the weak-coupling limit  \cite{Jankovic_2024}. That is why the DMFT solution for conductivity is almost exact in the weak-coupling limit.

In the QC solution the electron dynamics is obtained from a solution of the Schr\" odinger equation, while the ion dynamics is treated classically \cite{Tully_2023,Troisi_PRL2006,Fetherolf_PRX2020}. The electron Hamiltonian is given by
\begin{equation}\label{eq:QC}
 H^{\mathrm{el}} = -t_0 \sum_i \left( c_i^\dagger c_{i+1} + \mathrm{h.c.} \right) -g\sqrt{2\omega_0}\sum_i x_i c_i^\dagger c_i,
\end{equation}
where the displacement operator 
$x_i = (1/\sqrt{2\omega_0}) (a_i^\dagger + a_i) $ 
is considered as a classical variable. The ion dynamics $x(t)$ is treated both within the classical path approximation (CPA) and the mean-field Ehrenfest method \cite{Fetherolf_PRX2020,Runeson_2024}. In CPA, the ions perform harmonic oscillations, $x_i(t) = x_i(0)\cos(\omega_0 t) + (\dot x_i(0)/\omega_0) \sin(\omega_0 t)$. Hence, in this case the ion dynamics is completely determined by the initial ion displacements and velocities. As we show in Supplemental Material (SM) Sec.~I (see also references \cite{Cohen_Tannoudji,Greiner_book,Landau_QMbook,Leeuwen_Notes2005,Aoki_2014,Wang_2011} therein), $x_i(0)$ and $\dot x_i(0)$ should be taken from the Gaussian distributions with the variance $\langle x_i^2 (0) \rangle = \frac{1}{2\omega_0} \coth \left( {\beta \omega_0} / {2}\right)$ and $\langle \dot x_i^2(0) \rangle = \frac{\omega_0}{2} \coth \left( {\beta \omega_0} / {2}\right)$, respectively. In the mean-field Ehrenfest method the ion dynamics is modified by the presence of electron by the back-action term $- \partial / \partial x_i \langle \langle \psi_n(t)| H^{\mathrm{el}}|\psi_n(t) \rangle \rangle$, where the outer angle bracket denotes averaging over the Boltzmann factor. The initial electron wave functions $\psi_n(0)$ correspond to the eigenstates of $H^{\mathrm{el}}$ with energy $E_n$ obtained for random ion displacements and velocities, $H^{\mathrm{el}}(0) |\psi_n(0)\rangle = E_n |\psi_n(0)\rangle$. Then we use the $4^{\mathrm{th}}$ order Runge-Kutta method for CPA or $2^{\mathrm{nd}}$ order Verlet method for Ehrenfest dynamics, using sufficiently small time step $\Delta t$, to calculate $\psi_n(t)$ from the time-dependent Schr\"odinger equation $i\frac{\partial}{\partial t} |\psi_n(t)\rangle = H^{\mathrm{el}}(t) |\psi_n(t)\rangle$. The current-current correlation function is then given by
\begin{equation}
 C_{jj}(t) =  \frac{1}{Z} \sum_{n,m} e^{-\beta E_n} \langle \psi_n(t)| j |\psi_m(t)\rangle \langle \psi_m| j |\psi_n \rangle ,
\end{equation}
which needs to be averaged over many realizations of initial ion positions and velocities. The QC equations are described in more detail in SM Sec.~\ref{SM:prva_glava}.

{\it Results.} We will present the results for $\omega_0=1/3$ at intermediate  ($\lambda=0.5$) and weak ($\lambda=1/8$) electron-phonon coupling, and the results for $\omega_0=0.1$ at $\lambda=0.45$ and $1.25$.

\begin{figure}[t]
\includegraphics[width=\columnwidth]{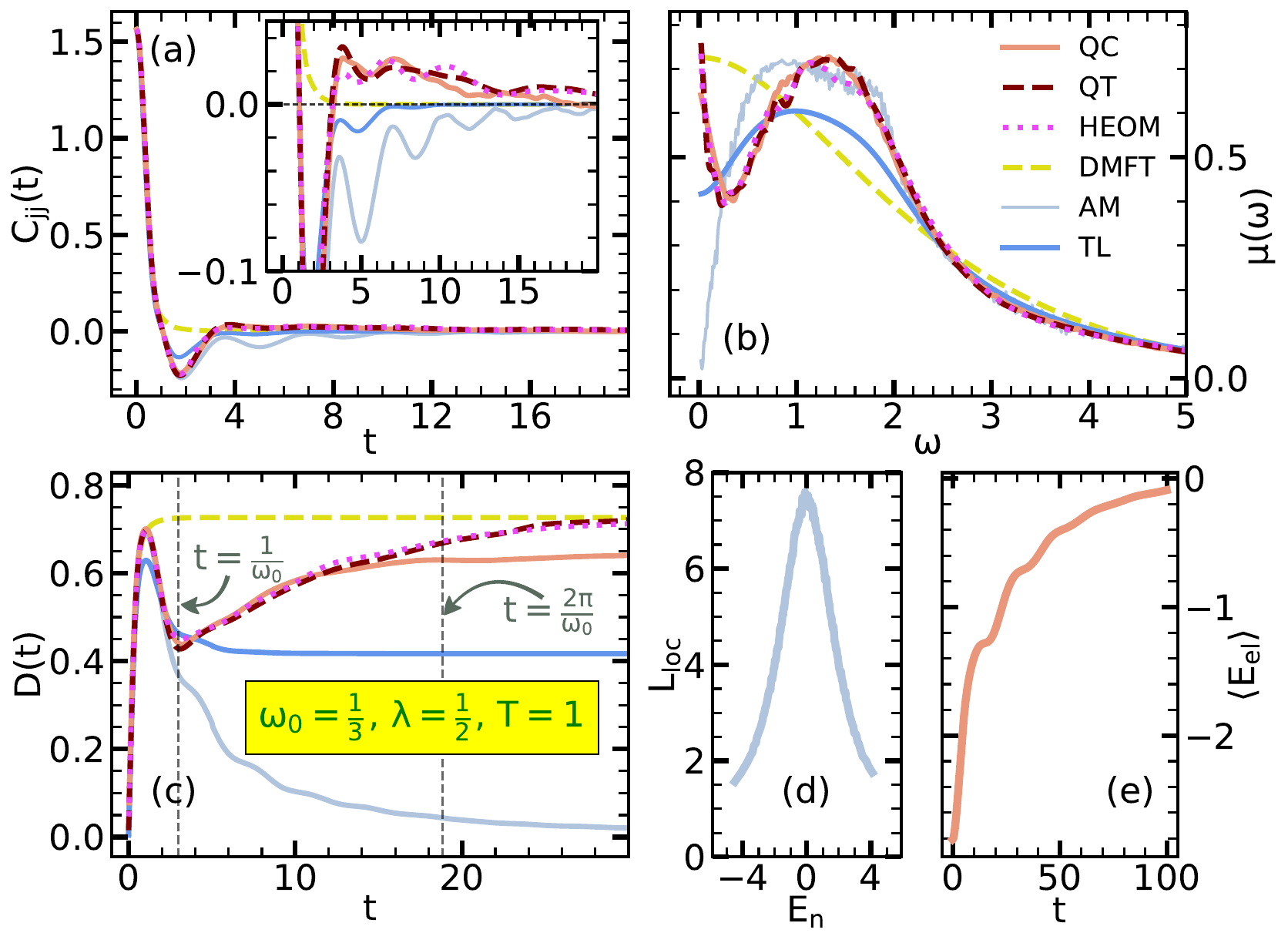}
\caption{Comparison between different methods of the
\begin{minipage}[t]{0.49\textwidth}
\justifying $\!\!\!\!\!\!$% Requires \usepackage{ragged2e}
(a) current-current correlation function, (b) frequency-dependent mobility, (c) time-dependent diffusion constant, (d) localizaton length and (e) ensemble-averaged electron kinetic energy for intermediate electron-phonon coupling.
        \end{minipage}
    }
%\caption{
%Comparison between different methods of the (a) current-current correlation function, (b) frequency-%dependent mobility, (c) time-dependent diffusion constant, (d) localizaton length and (e) ensemble-averaged %electron kinetic energy for intermediate electron-phonon coupling.
%}
\label{Fig1}
\vspace*{-0.5cm}
\end{figure}

{\it A.} The results for $\omega_0=1/3, \lambda=0.5$, ($g=0.577$) and $T=1$ are shown in Fig.~\ref{Fig1}. We start our analysis from the two well-known limits. The DMFT solution \cite{Mitric_2022,Ciuchi1997} neglects the vertex corrections to conductivity. In this case the current-current correlation function $C_{\mathrm{jj}}^{\mathrm{DMFT}}(t)$ exponentially goes to zero [Fig.~\ref{Fig1}(a)] and the frequency-dependent mobility $\mu(\omega)$ assumes a Lorentzian shape [Fig.~\ref{Fig1}(b)]. Since the self-energy in the Holstein model is almost local \cite{Mitric_2022}, the DMFT solution practically coincides with the full quantum solution in the bubble approximation \cite{Jankovic_2024}. In the static case, when the ion vibrations are frozen at their randomly chosen $t=0$ positions, we end up with the AM solution. In this case, $C_{jj}^{\mathrm{AM}}(t)$ changes sign and then, following a power law $\propto 1/t^2$, goes to zero, such that the dc mobility $\mu_{\mathrm{dc}} = \int_0^\infty C_{jj}^{\mathrm{AM}}(t) = 0$.

The qualitative difference which brings the full solution of the Holstein model is best understood by looking at the time-dependent diffusion constant $D(t)$ [Fig.~\ref{Fig1}(c)]. After the ballistic regime where $D(t)$ increases linearly, there is a plateau in the DMFT solution corresponding to the diffusive transport. In the AM $D(t)$ reaches a maximum at time $t_{\mathrm{max}}$ and then decreases towards zero. $t_{\mathrm{max}}$ increases with the increase of the localization length. The full solution also features a decrease in diffusion which is, however, interrupted at time $t \approx 1/ \omega_0$ when $D(t)$ starts to increase again.
This corresponds to the small positive values of $C_{jj}(t)$ for $1/\omega_0 \lesssim t \lesssim 2\pi/\omega_0$, see the inset of Fig.~\ref{Fig1}(a).  The increase in $D(t)$ causes the appearance of an additional zero-frequency peak in $\mu(\omega)$ [Fig.~\ref{Fig1}(b)]. The agreement between the QC and quantum QT and HEOM solutions is excellent up to $t \lesssim 2\pi/\omega_0$. For larger times, $D(t)$ reaches a clear plateau in QC, while there is further slight increase in QT and HEOM solutions. We cannot say if this is due to finite size of the lattice in a quantum solution ($N=7$ in QT and $N=10$ in HEOM). We note that the QC solution is here obtained on the lattice with $N=200$ sites after averaging over 3000 initial ion displacements and velocities. The back-action term within the Ehrenfest approach leads to very small differences and we show just the CPA results in main text. For details of the QC numerics see SM Sec.~\ref{SM:glava_druga}. 

\begin{figure}[t]
\includegraphics[width=\columnwidth]{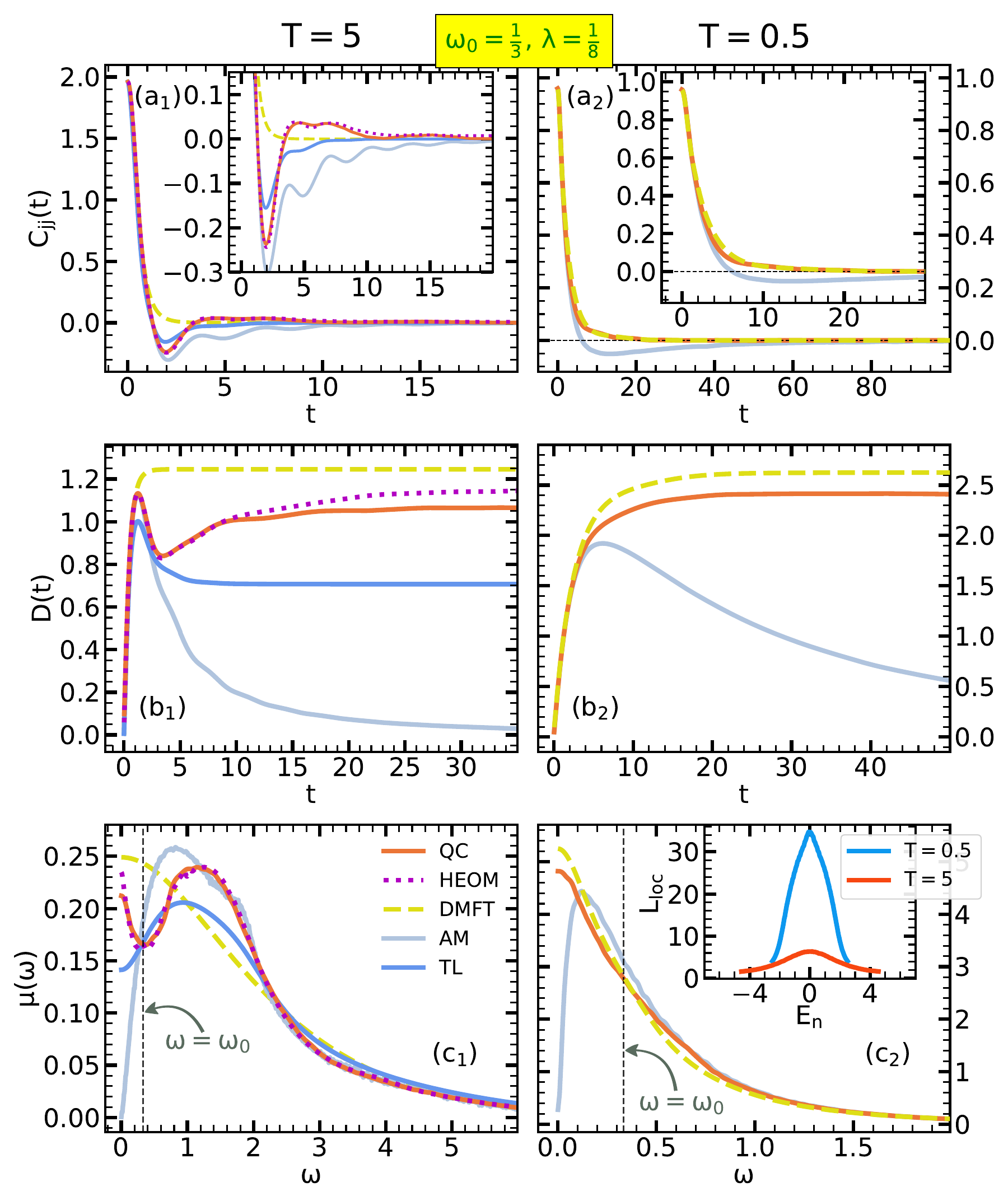}
\caption{(a) Current-current correlation function, (b) time-
\begin{minipage}[t]{0.49\textwidth}
\justifying $\!\!\!\!\!\!$% Requires \usepackage{ragged2e}
dependent diffusion constant and (c) frequency-dependent mobility  for weak electron-phonon coupling at high (left column) and low temperature (right column).
        \end{minipage}
    }

\label{Fig2}
\vspace*{-0.5cm}
\end{figure}

We now discuss few conceptual issues. First, we note that for the appearance of both the maximum and the minimum in $D(t)$ we need a well-separated time scales: $t_{\mathrm{max}}$ being smaller than $1/ \omega_0$. $t_{\mathrm{max}}$ is proportional to the localization length $L_{\mathrm{loc}}$ in the static case ($L_{\mathrm{loc}}(E_n) =1/ \sum_i |\psi_n^i|^4$, where $\psi_n^i$ are the components of $|\psi_n(0)\rangle$) and at $ t \sim 1/ \omega_0$ the inelastic electron-phonon scattering comes into effect. The localization length [Fig.~\ref{Fig1}(d)] depends on the eigenstate energy $E_n$, while the states near the lower band edge participate in charge transport at low and intermediate temperatures.
Notion of these two time scales forms the basis of popular transient localization scenario of charge transport introduced by Ciuchi, Fratini and Mayou \cite{Ciuchi_PRB2011,Fratini2016}. 
Yet, the phenomenological TL approach cannot explain the upturn in $\mu(\omega)$ for $\omega < \omega_0$. Within TL, the correlation function 
$C_{jj}^{\mathrm{TL}}(t)=C_{jj}^{\mathrm{AM}}(t)e^{-t/\tau_{\mathrm{in}} }$ exponentially goes to zero, and $D(t)$ just goes to a constant for $t \gtrsim 1/\omega_0$. We note that the increase of the ensemble-averaged electron kinetic energy $\langle E_{\mathrm{el}}(t)\rangle$ [Fig.~\ref{Fig1}(e)], which is a well-known artifact of QC dynamics \cite{Fratini2016}, does not significantly influence $C_{jj}^{\mathrm{QC}}(t)$ since the timescale of this increase is significantly longer than the time it takes for $C_{jj}$ to decay to zero. This is in agreement with the findings from a very recent QC study on the Peierls model \cite{Runeson_2024}. 
Finally, by taking the correlation functions from Fig.~\ref{Fig1} as an example, in SM Sec.~\ref{SM:glava_druga}  we demonstrate why it is impossible to reliably extract the dc mobility just from the imaginary-axis data, which one could obtain from the QMC calculations.

{\it B.} We examine the influence of the temperature to charge transport in Fig.~\ref{Fig2}. We set weaker electron-phonon coupling $\lambda=1/8$ ($g=0.288$) and consider $T=5$ (left column) and 0.5 (right column). At high temperature [Fig.~\ref{Fig2}(a$_1$)-(c$_1$)], the localizaton length is small [inset of Fig.~\ref{Fig2}(c$_2$)], and  the charge dynamics is qualitatively the same as in Fig.~\ref{Fig1}: Apart from the DDP at finite frequency, there is an additional zero-frequency peak in $\mu(\omega)$. The QC solution (for $N=100$) agrees very well with the quantum (HEOM) solution (for $N=7$) \cite{Jankovic_2023,Jankovic_2024}. At lower temperature [Fig.~\ref{Fig2}(a$_2$)-(c$_2$)], one needs longer chain to eliminate the finite-size effects and the HEOM or QT solution are not available. Yet, for a comparison we can use the DMFT solution, which is in the thermodynamic limit, since we know that the importance of vertex corrections decreases for weaker electron-phonon interaction and lower temperature \cite{Jankovic_2024}. DMFT and QC results are in excellent agreement. This is favored by slower increase of $\langle E_{\mathrm{el}}(t)\rangle$ in the weak-coupling case, see SM Sec.~\ref{SM:glava_druga}. The dynamical disorder is small at $T=0.5$, which corresponds to large localization length and $t_{\mathrm{max}}$. Since $t_{\mathrm{max}} > 1 / \omega_0$, here we do not observe the TL phenomenology.

\begin{figure}[t]
\includegraphics[width=\columnwidth]{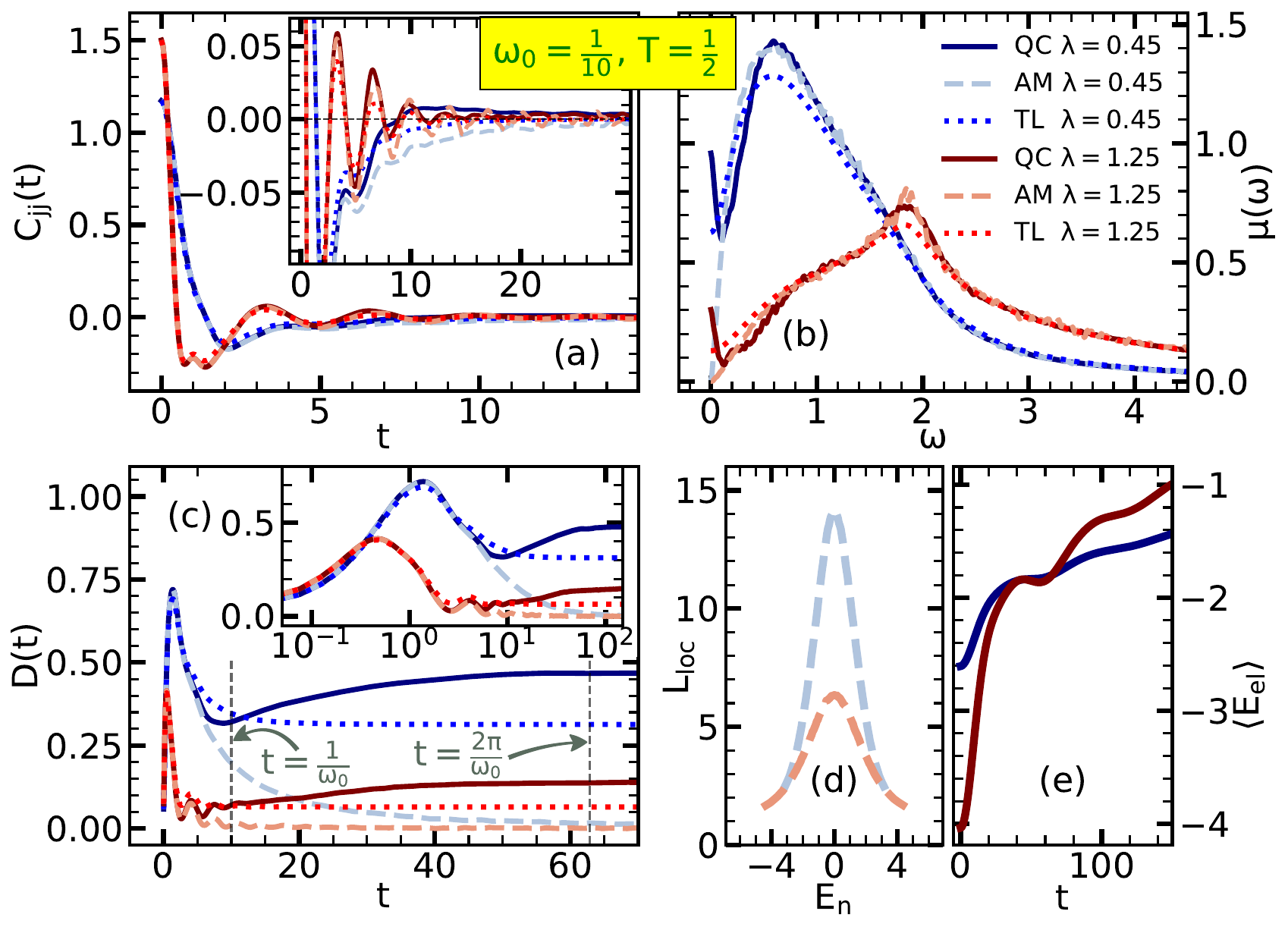}
\caption{The same quantities as in Fig.~\ref{Fig1} for lower phonon
\begin{minipage}[t]{0.49\textwidth}
\justifying $\!\!\!\!\!\!$% Requires \usepackage{ragged2e}
frequency and for an intermediate and strong electron-phonon coupling.
        \end{minipage}
    }

\label{Fig3}
\vspace*{-0.5cm}
\end{figure}

{\it C.} The electron dynamics for lower phonon frequency $\omega_0=0.1$ is presented in Fig.~\ref{Fig3}. Here we do not have a fully quantum solution since one would need a large lattice-size and propagation of the correlation function up to very long times. Yet, the previous analysis gives us confidence that the QC approach gives a proper description of the electron dynamics also for lower phonon frequency. The results at $T=0.5$ for $\lambda=0.45$ ($g=0.3$) and $\lambda = 1.25$ ($g=0.5$) look qualitatively the same as in Fig.~\ref{Fig1}. Most importantly, in the time interval $1/\omega_0 \lesssim t \lesssim 2\pi/\omega_0$ the correlation function assumes small positive values
[Fig.~\ref{Fig3}(a)], which leads to the upturn in $\mu(\omega)$ for $\omega < \omega_0$ [Fig.~\ref{Fig3}(b)]. Diffusion constant $D(t)$ reaches a plateau at $t \sim 2\pi / \omega_0$, as before [Fig.~\ref{Fig3}(c)]. The displaced peak in $\mu(\omega)$ is centered at lower frequency for $\lambda=0.45$ than for $\lambda=1.25$ when in the AM the localization length is longer [Fig.~\ref{Fig3}(d)]. We expect that the QC solution is quantitatively better for weaker electron-phonon coupling, when the increase in the electron kinetic energy in QC solution is not substantial by the time that the plateau in $D(t)$ is reached [Fig.~\ref{Fig3}(e)]. The TL solution, which does not feature the upturn at long times (low frequency), is shown for comparison.

%We can define a characteristic localization length, $L_{\mathrm{loc}}^{\mathrm{ch}} = \sum_{E_n} L_{\mathrm{loc}}(E_n) e^{-\beta E_n} / \sum_{E_n} e^{-\beta E_n}$, as a measure of the localization of an electron participating into the transport. $L_{\mathrm{loc}}^{\mathrm{ch}} (g=0.5) = $

{\it Conclusions.}
In summary, we performed the QC calculations of charge transport in the 1D Holstein model for two representative phonon frequencies. Frequency $\omega_0 = 0.1$ for temperature $T=0.5$ is appropriate for modelling of organic semiconductors like rubrene, and the results for $\omega_0 = 1/3$ and various $g$ and $T$ are used for a comparison with fully quantum HEOM \cite{Jankovic_2023,Jankovic_Zenodo8} and QT solutions \cite{Mitric_preprint}. The agreement between these two quantum methods is excellent. Both methods can propagate the real-time correlation function up to long times and for lattice sizes which are representative of the thermodynamic limit. These properties gave us an opportunity to make detailed comparisons with the QC dynamics, which is performed on much longer chains, consisting of few hundred lattice sites. For temperatures $T \gtrsim \omega_0$, we find that the charge dynamics is qualitatively the same, and quantitatively quite similar to the quantum case. For strong dynamical disorder, we observe a non-monotonic diffusion $D(t)$. The maximum in $D(t)$ at short times correspond to the displaced Drude peak in $\mu(\omega)$ as a precursor to the Anderson localization. At times $1/\omega_0 \lesssim t \lesssim 2\pi/\omega_0 $, diffusion increases before reaching a plateau at $t \sim 2\pi / \omega_0$. This corresponds to the upturn in $\mu(\omega)$ for $\omega < \omega_0$. This feature appears regardless of the specific methodology that we used, which gives us confidence that it is a genuine property of the model.
The zero-frequency peak in $\mu(\omega)$ does not appear in the TL model of charge transport \cite{Fratini2016} which, therefore, underestimates the dc mobility. Such peak is also observed in very recent QC calculations in the Peierls model \cite{Runeson_2024}. This indicates that the peak appears both for local and nonlocal electron-phonon coupling.
Our work overcomes the constraints of a small lattice size in real-frequency Lanczos calculations or analytical continuation of the imaginary-frequency data \cite{Buividovich_2024} and, to our knowledge, presents a first explicit comparison between the charge transport in models featuring quantum and classical phonons.

We thank V.~Jankovi\'c for fruitful discussions. D.~T. thanks J.~Mravlje for useful discussions. 
P.~M.~and D.~T.~acknowledge funding
provided by the Institute of Physics Belgrade through
a grant from the Ministry of Science, Technological Development, and Innovation of the Republic of Serbia.
Work in Florida (V.D.) was supported by the NSF Grant No. DMR-2409911, and the National High Magnetic Field Laboratory through the NSF Cooperative Agreement No. DMR-2128556 and the State of Florida.

{\it Data availability.} The data that support the findings of this paper are openly available \cite{zenodo_data}. 

%\input{Appendix}
%\bibliographystyle{apsrev4-1}
%\bibliography{refs_TL.bib}


%%%%%%%%%%%%%%%%%%%%%%%%%%%%%%%%%%%%%%%%%%%
%%%%%%%%%%%%%%%%%%%%%%%%%%%%%%%%%%%%%%%%%%%
%%%%%%%%%%%%%%%%%%%%%%%%%%%%%%%%%%%%%%%%%%%
%%%%%%%%%%%%%%%%%%%%%%%%%%%%%%%%%%%%%%%%%%%
%%%%%%%%%%%%%%%%%%%%%%%%%%%%%%%%%%%%%%%%%%%
%%%%%%%%%%%%%%%%%%%%%%%%%%%%%%%%%%%%%%%%%%%

\clearpage
\pagebreak
\newpage

%
%\begin{widetext}
%\begin{center}
%\textbf{\large Supplemental Material}
%\end{center}
%\end{widetext}

%\begin{center}
%\textbf{\large Supplemental Material}
%\end{center}

\onecolumngrid
\begin{center}
  \textbf{\large Supplemental Material \\ Precursors to Anderson Localization in the Holstein Model: Quantum and
Quantum-Classical Solutions}\\[.2cm]
  P. Mitri\'c,$^{1}$ V. Dobrosavljevi\'c,$^{2}$ and D. Tanaskovi\'c$^1$\\[.1cm]
  {\itshape ${}^1$Institute of Physics Belgrade, University of Belgrade, Pregrevica 118, 11080 Belgrade, Serbia} \\[.1cm]
  {\itshape ${}^2$Department of Physics and National High Magnetic Field Laboratory, \\ Florida State University, Tallahassee, Florida, USA}
  \\[1cm]
\end{center}
\twocolumngrid

\setcounter{equation}{0}
\setcounter{figure}{0}
\setcounter{table}{0}
\makeatletter
\renewcommand{\theequation}{S\arabic{equation}}
\renewcommand{\thefigure}{S\arabic{figure}}
\renewcommand{\bibnumfmt}[1]{[S#1]}
\renewcommand{\citenumfont}[1]{S#1}
\renewcommand{\thetable}{S\arabic{table}}

\section{Quantum-classical dynamics: Formalism} \label{SM:prva_glava}

This Section of the Supplemental Material is presented for clarity and pedagogical reasons. We follow the quantum-classical (QC) approach from Refs.~\cite{SMTroisi_PRL2006,SMCiuchi_PRB2011,SMFetherolf_PRX2020}.

\subsection{Probability distribution for the phonon coordinate}

%Here we derive the probability distribution function of the phonon coordinate for a quantum harmonic oscillator.
%This results is known and it is used in quantum-classical calculations as in Refs.~\cite{Troisi_PRL2006,Ciuchi_PRB2011,Fetherolf_PRX2020}. 
%The derivation follows the Cohen-Tannoudji book [x].

We consider just the phonon part of the Hamiltonian since the electron concentration goes to zero.
The Hamiltonian is given by
\begin{equation}
H = \omega_0 a^\dagger a .
\end{equation}
By definition, the probability distribution function is given by
\begin{eqnarray} \label{eq:p_x}
p(x) &=& \frac{1}{Z} \sum_{n=0}^\infty e^{-\beta n \omega_0} 
|\langle x | \psi_n \rangle|^2 \nonumber \\
&=& 
\frac{1}{Z}  
\langle x | \underbrace{\sum_{n=0}^\infty  e^{-\beta n \omega_0} |\psi_n \rangle \langle \psi_n}_{\equiv e^{-\beta H}} | x \rangle \nonumber \\
&=& 
\frac{1}{Z} \langle x | e^{-\beta H} | x\rangle,
\end{eqnarray}
where $\beta = \frac{1}{T}$, Z is the partition function and $\psi_n$ are the eigenstates of Hamiltonian (1), i.e. the Hermite functions. The expression on the right hand side of Eq.~(\ref{eq:p_x}) corresponds to the density matrix of a harmonic oscillator expressed in the coordinate representation. The explicit expression for this quantity can be found in many textbooks \cite{SMCohen_Tannoudji,SMGreiner_book,SMLandau_QMbook}, but for the sake of clarity and completeness, we will here repeat that derivation following the approach outlined by \cite{SMCohen_Tannoudji}.

%\subsection{Calculating $p(x)$}

%From now on, the term $\frac{1}{Z}$ will be dropped - this is justified as long as we normalize $p(x)$ at the end of the calculation. Since $p(x)$ is no longer normalized, it does not provide any useful information for a single value of $x$. This is why we want to examine its value when we vary the argument, i.e. $p(x+dx)$. It can be expressed using the momentum operator $P$ as follows
Since $p(x)$ will be normalized at the end of the calculation, the term $1/Z$ will be omitted in $p(x)$ from this point onward. Without normalization, $p(x)$ no longer holds meaningful information for a single value of $x$. Instead, our focus shifts to examining how $p(x)$ behaves as its argument varies. It can be expressed using the momentum operator $P$ as follows
\begin{eqnarray} \label{Eq:p_eq}
p(x+dx) &=& \langle x+dx | e^{-\beta H} | x+dx \rangle \nonumber \\
&=& p(x) 
+ i dx \langle x | \left[ P, e^{\beta H} \right] | x \rangle.
\end{eqnarray}
This expression is a consequence of the fact that the momentum operator P is the generator of translations
\begin{equation}
| x + dx \rangle = e^{-i dx P} | x \rangle \approx 
\left( 1 - i dx P \right)
| x \rangle.
\end{equation}
The second term in Eq.~(\ref{Eq:p_eq}) could be evaluated immediately if (some function) of the coordinate operator $X$ was there instead of the momentum operator. This holds true because $X$ acts trivially on $|x\rangle$. Such a transformation actually does exist and is given by the following theorem
%Our task is thus reduced to the calculation of the second term in Eq.~\eqref{Eq:p_eq}. This could be easily accomplished if there was the coordinate operator $X$ instead of the momentum operator in Eq.~\eqref{Eq:p_eq} (since $X$ acts trivially on $| x \rangle$). In that sense, the following theorem practically solves our problem.
%

\vspace*{.2cm}

{\bf{Theorem:}}
The following relation always holds
\begin{equation}
\left[ 
P, e^{-\beta H}
\right] = i\omega_0 
\{ X, e^{-\beta H} \} 
\; \mathrm{th} \left( \frac{\beta \omega_0}{2} \right),
\end{equation}
where $\{, \}$ is the anticommutator.
%
%\end{theorem}
%

%\begin{proof}

\vspace*{.2cm}
{\it Proof.}
Starting from
%$P \propto a-a^\dagger$, it is useful to examine what happens if we try to commute $a$ (and $a^\dagger$) with $e^{-\beta H} = e^{-\beta \omega_0 a^\dagger a} $. This is facilitated using
the Baker--Campbell--Hausdorff formula
\begin{equation}
e^B A e^{-B} = A + [B,A] + \frac{1}{2!} [B, [B, A]] + \dots.
\end{equation}
and setting $B= -\beta \omega_0 a^\dagger a$, $A = a$, one finds that
%One finds
\begin{eqnarray}
e^{-\beta \omega_0 a^\dagger a} a e^{\beta\omega_0 a^\dagger a} \! \! &=& \! \!
a - \beta \omega_0 [ a^\dagger a, a] \nonumber \\ 
&+& \! \! \frac{\beta^2 \omega_0^2}{2!}
[a^\dagger a, [a^\dagger a, a ]] + \dots = a e^{\beta \omega_0},
\end{eqnarray}
and as a consequence
%Hence
%
\begin{subequations} 
\begin{equation} \label{Eq:a1_transf}
 e^{-\beta \omega_0 a^\dagger a} a  = e^{\beta\omega_0} a e^{-\beta\omega_0 a^\dagger a},
\end{equation}
%
%\vspace*{-.5cm}
%and analogously
\begin{equation} \label{Eq:a2_transf}
 e^{-\beta \omega_0 a^\dagger a} a^\dagger = e^{-\beta\omega_0} a^\dagger e^{-\beta\omega_0 a^\dagger a}.
\end{equation}
\end{subequations}
The momentum $P\sim a-a^\dagger$ and coordinate $X\sim a+a^\dagger$ operator will pop out if we rewrite the exponential $e^{-\beta \omega_0}$ in Eqs. (\ref{Eq:a1_transf}) and (\ref{Eq:a2_transf}) as
%We would now like to add/subtract these two equations and somehow get the terms proportional to $a-a^\dagger$, giving momentum operator and terms proportional to $a+a^\dagger$, giving coordinate operator. This can be accomplished by writing the exponential $e^{-\beta \omega_0}$ as
%
\begin{equation}
e^{\beta \omega_0} = \frac{1+x}{1-x} \implies x = 
\mathrm{th} \left( 
\frac{\beta\omega_0}{2}
\right)
\end{equation}
Eqs.~\eqref{Eq:a1_transf}~and~\eqref{Eq:a2_transf} become
\begin{subequations}
\begin{eqnarray} \label{Eq:a_trans1_th}
\left[ 1 - \mathrm{th} \left( \frac{\beta\omega_0}{2} \right) \right] 
 e^{-\beta\omega_0 a^\dagger a} \; a  \nonumber \\ =
\left[ 1 + \mathrm{th} \left( \frac{\beta\omega_0}{2} \right) \right]
 a \; e^{-\beta\omega_0 a^\dagger a} , \\
 \left[ 1 + \mathrm{th} \left( \frac{\beta\omega_0}{2} \right) \right]
e^{-\beta\omega_0 a^\dagger a} \; a^\dagger \nonumber \\ =
\left[ 1 - \mathrm{th} \left( \frac{\beta\omega_0}{2} \right) \right]
 a^\dagger \; e^{-\beta\omega_0 a^\dagger a} . \label{Eq:a_trans2_th}
\end{eqnarray}
\end{subequations}
Subtracting Eq.~\eqref{Eq:a_trans2_th} from Eq.~\eqref{Eq:a_trans1_th} and simplifying the obtained expression, we get
\begin{equation}
[a- a^\dagger, e^{-\beta H}] = - \{ a+a^\dagger, e^{-\beta H} \} \mathrm{th} \left( \frac{\beta \omega_0}{2} \right).
\end{equation}
By multiplying both sides with $-i \sqrt{\frac{\omega_0}{2}}$  we arrive at the final expression
%we finally obtain
%
\begin{equation}
\left[ 
P, e^{-\beta H}
\right] = i\omega_0 
\{ X, e^{-\beta H} \} 
\; \mathrm{th} \left( \frac{\beta \omega_0}{2} \right),
\end{equation}
which proves the theorem. $\square$
%\end{proof}

\vspace*{.2cm}

Plugging this back into Eq.~\eqref{Eq:p_eq}
%Let us now use this theorem in Eq.~\eqref{Eq:p_eq}
\begin{eqnarray}
\frac{dp(x)}{dx} &=& -  \omega_0 \mathrm{th} \left( \frac{\beta\omega_0}{2} \right) \langle x | \{X, e^{-\beta H} \} | x \rangle \nonumber \\
&=& 
-2  \omega_0 \mathrm{th} \left( \frac{\beta\omega_0}{2} \right)  x\, p(x) ,
\end{eqnarray}
%
%This is a simple
we arrive at a well known differential equation whose solution is the Gaussian
\begin{equation} \label{Eq:X_distribution}
p(x) = \frac{1}{\sigma \sqrt{2\pi}} e^{-\frac{x^2}{2\sigma^2}}; \qquad 
\sigma^2 = \frac{1}{2\omega_0} \mathrm{coth}
\left( \frac{\beta\omega_0}{2} \right) .
\end{equation}
%

%\subsection{The site disorder in the Holstein model}

Since the on-site interaction in the Holstein model is of the form
\begin{equation}
H_{int} = -g (a + a^\dagger) n = -g \sqrt{2\omega_0} X n,
\end{equation}
where $n = c^\dagger c$, , we see that it can be rewritten as
%This interaction part of the Hamiltonian can now be substituted with
%
\begin{equation}
H_{int} \to \varepsilon n; \qquad \varepsilon = -g \sqrt{2\omega_0} X ,
\end{equation}
where, $\varepsilon$ can also be regarded as random variable.
%with the appropriate distribution.
As a consequence of the fact that the probability distribution of $X$ follows a Gaussian (see Eq.~\eqref{Eq:X_distribution}), it can be concluded that $\varepsilon$ will likewise exhibit a Gaussian distribution, centered at zero, with the variance given by
%Since the distribution of $X$ is given by the Gaussian in Eq.~\eqref{Eq:X_distribution}, we conclude that $\varepsilon$ will also have a Gaussian distribution, centered around zero, with the following variance
%
\begin{eqnarray}\label{eq:variance}
\sigma_{\varepsilon}^2 &\equiv& \mathrm{Var} \left[ \varepsilon \right] = 2\omega_0 g^2 \mathrm{Var} \left[ X \right] = g^2  \mathrm{coth}
\left( \frac{\beta\omega_0}{2} \right) \nonumber \\ &=& 2g^2 \left(
\frac{1}{2} + \frac{1}{e^{\beta\omega_0} -1}
\right).
\end{eqnarray}
Therefore, the probability distribution for $\varepsilon$ reads as follows
%Hence, the probability distribution for $\varepsilon$ is given by
%
\begin{equation}
p_{\varepsilon}(\varepsilon) = \frac{1}{\sigma_{\varepsilon} \sqrt{2\pi}} e^{-\frac{\varepsilon^2}{2\sigma_{\varepsilon}^2}} .
\end{equation}

\subsection{Kubo formula for a time-dependent Hamiltonian}\label{Sec: Kubo}

Our goal is to calculate the current-current correlation function for an electron which scatters from classical vibrations using the Kubo formula for the time-dependent Hamiltonian. Other quantities, like the time-dependent diffusion constant and frequency-dependent mobility, can be directly obtained from the current-current correlation function.
%We want to make a comparison with the Holstein model with quantized phonons and, in addition, to explore the finite size effect within the semiclassical approach.
%
Here we derive the Kubo formula in the case when the unperturbed (without the external field) Hamiltonian $H_0$ is time dependent. 

The total Hamiltonian is given by
\begin{equation} \label{H_total}
 H(t) = H_0(t) + H'(t),
\end{equation}
and we look for the linear response to $H'$. Following Refs.~\cite{SMAoki_2014,SMLeeuwen_Notes2005}, we assume that at $t_0=0$ the electron is in thermodynamic equilibrium corresponding to a given static ion displacements, described by the Hamiltonian $H_0(0)$ and the partition function $Z=\mathrm{Tr}(e^{-\beta H_0(0)})={\mathrm{Tr}} (\rho(0))$, with the eigenstates denoted by $|n\rangle$. Then, at $t_0=0$ we allow both the classical ion vibrations (leading to the time-dependent Hamiltonian $H_0(t)$) and the external field (described by the Hamiltonian $H'(t))$. The eigenstates evolve with time according to
\begin{equation}\label{n_S}
 |n(t)\rangle = U(t,t_0)|n(t_0)\rangle ,
\end{equation}
where the time evolution operator $U$ is determined by
\begin{equation}
 i\partial_t U(t,t_0) = H(t) U(t,t_0) .
\end{equation}
The expectation value of an operator $A$ is given by
\begin{equation}\label{A_H}
 \langle A \rangle (t) = \frac{1}{Z} \sum_n \langle n(t)| A  |n(t)\rangle e^{-\beta E_n} = {\mathrm{Tr}}(\rho(0)A(t)),
\end{equation}
where $A(t) = U(t_0,t)AU(t,t_0)$. Alternatively, we can write that

\begin{equation}\label{A_H_2}
 \langle A \rangle (t) = {\mathrm{Tr}}(\rho(t)A),
\end{equation}
where $\rho(t)= U(t,t_0)\rho(0)U(t_0,t)$ is the time-dependent density matrix.

We closely follow the derivation of the Kubo formula in the standard case when the Hamiltonian $H_0$ is time-independent \cite{SMMahan}. We first need to define the interaction picture (with respect to $H_0(t)$), which is done by the following relation
\begin{equation}\label{interaction_def}
 |\hat{n}(t)\rangle = U_0^\dagger(t,t_0)|n(t)\rangle ,
\end{equation}
where
\begin{equation}
 i\partial_t U_0(t,t_0) = H_0(t) U_0(t,t_0) .
\end{equation}
The evolution operator in the interaction picture is defined by the relation
\begin{equation}
 |\hat{n}(t)\rangle = \hat{U}(t,t_0)|\hat{n}(t_0)\rangle ,
\end{equation}
All the quantities in the interaction picture will have a caret sign. Note that in the stationary case $U_0(t,t_0) = e^{-iH_0(t-t_0)}$. It remains to prove that
\begin{equation}
 i\partial_t \hat{U}(t,t_0) = \hat{H'}(t) \hat{U}(t,t_0) ,
\end{equation}
where $\hat{H'}(t) = U_0(t_0,t) H'(t)U_0(t,t_0)$.

\vspace*{.2cm}

{\it Proof.} From Eqs.~\ref{n_S} and \ref{interaction_def} $|\hat{n}(t)\rangle = U_0^\dagger(t,t_0)|n(t)\rangle = U_0^\dagger(t,t_0)U(t,t_0) |n(t_0)\rangle $. Therefore, $\hat{U}(t,t_0) = U_0^\dagger(t,t_0)U(t,t_0) $. Then,
\begin{eqnarray}
 i \partial_t \hat{U}(t,t_0)= i \partial_t \left( U_0^\dagger(t,t_0)U(t,t_0) \right) \nonumber \\
 = \left( i \partial_t U_0^\dagger(t,t_0) \right) U(t,t_0) + U_0^\dagger(t,t_0)\left( i \partial_t U(t,t_0) \right)  \nonumber \\
= -\left( H_0(t) U_0(t,t_0)\right)^\dagger U(t,t_0) + U_0^\dagger(t,t_0)H(t)U(t,t_0) \nonumber \\
= - U_0(t,t_0)^\dagger H_0(t) U(t,t_0) + U_0^\dagger(t,t_0)H(t)U(t,t_0) \nonumber \\
= U_0(t,t_0)^\dagger \left( H(t)-H_0(t)\right) U(t,t_0) \nonumber \\
= U_0(t,t_0)^\dagger H'(t) U_0(t,t_0) \hat{U}(t,t_0) \nonumber \\
= \hat{H}'(t) \hat{U}(t,t_0) . \, \, \, \, \square \nonumber
\end{eqnarray}

We are now ready to derive the Kubo formula for $\langle A \rangle (t)$. For now, we assume that $A$ is time-independent in the Schr\"odinger picture.
We will use that $\hat{U}(t,t_0) \approx 1-i\int_{t_0}^t dt' \hat{H}'(t')$. We find
\begin{eqnarray}
\langle A \rangle (t) =  \frac{1}{Z} \sum_n \langle n(t)| A  |n(t)\rangle e^{-\beta E_n} \nonumber \\
=  \frac{1}{Z} \sum_n \langle n| \hat{U}^\dagger(t,t_0) \hat{U}_0^\dagger(t,t_0)  A U_0(t,t_0)\hat{U}(t,t_0) |n\rangle e^{-\beta E_n} \nonumber \\
= \frac{1}{Z} \sum_n \langle n| \left( 1+i\int_{t_0}^t dt' \hat{H}'(t')\right) \hat{A}(t) \nonumber \\
\times \left( 1-i\int_{t_0}^t dt' \hat{H}'(t') \right) |n\rangle e^{-\beta E_n} \nonumber \\
= \langle \hat{A}(t) \rangle - i \int_{t_0}^t dt'\langle \left[ \hat{A}(t),\hat{H}'(t')\right] \rangle \,\,\,\,\,\, \, \,
\end{eqnarray}
The ensemble average is with respect to $H_0(t_0)$. We set $t_0 \rightarrow -\infty$ and   obtain
\begin{equation}
 \delta \langle A \rangle (t)  = \langle A \rangle (t) - \langle \hat{A} \rangle (t) = \int_{-\infty}^{\infty} dt' C_{AH'}^R(t,t') ,
\end{equation}
where the retarded response function is equal to
\begin{equation}
 C_{AH'}^R(t,t') = -i \theta(t-t') \langle \left[ \hat{A}(t),\hat{H}'(t')\right] \rangle
\end{equation}
Hence, everything looks the same as in the case with time-independent $H_0$. The only difference is that the ensemble average is taken with respect to $H_0(t=t_0)$ and in the time evolution $\hat{H'}(t) = U_0(t_0,t) H'(t)U_0(t,t_0)$ we cannot simply replace $U_0(t,t_0)$ by $e^{-iH_0(t-t_0)}$. We can also allow for an explicit time-dependence of $A(t)$ in the Schr\"odinger picture, as it would be needed for the Peierls model in the QC approximation.

The frequency-dependent mobility assumes the same form as for the time-independent $H_0$
\begin{equation}
  \mu(\omega) = \frac{1-e^{-\beta \omega}}{2N\omega} \int_{-\infty}^{\infty} dt e^{i\omega t} C_{jj}(t),
\end{equation}
where
\begin{eqnarray}
 C_{jj}(t) &=& {\mathrm{Tr}} \left[ j(t) j(0) e^{-\beta H_0(0)}\right] / Z \nonumber \\
 &=& \frac{1}{Z} \sum_n  e^{-\beta E_n}\langle n| j(t) j(0) |n\rangle ,
\end{eqnarray}
where $Z=\mathrm{Tr}(e^{-\beta H_0(0)})={\mathrm{Tr}} (\rho(0))$, $|n\rangle$ are eigenstates of $H_0(0)$ and $j(t)= U_0(0,t) j U_0(t,0)$.
We omit the caret sign from now on. In the Holstein model $j=i t_0 \sum_i \left( c_i^\dagger c_{i+1} - c^\dagger_{i+1} c_i \right)$. For numerical implementation we write
\begin{eqnarray}\label{eq_SM_Cjj}
 C_{jj}(t) &=& \frac{1}{Z} \sum_n \langle n| U_0(0,t) j U_0(t,0) j |n\rangle \nonumber \\
 &=& \frac{1}{Z} \sum_n \langle n| U_0^\dagger(t,0) j U_0(t,0) j |n\rangle \nonumber \\
 &=& \frac{1}{Z} \sum_{n,m} \left( U_0(t,0) |n\rangle \right)^\dagger j \left( U_0(t,0) |m\rangle \right) \langle m| je^{-\beta E_n}|n\rangle \nonumber \\
 &=& \frac{1}{Z} \sum_{n,m} e^{-\beta E_n} \langle n(t)| j |m(t)\rangle \langle m| j |n \rangle .
\end{eqnarray}

\subsection{Ehrenfest and CPA equations}

In the QC method, we treat the electron dynamics quantum-mechanically and the phonons are considered as classical vibrations. The electronic part of the Holstein Hamiltonian is given by

\begin{equation}\label{eq: H_el}
 H^{\mathrm{el}} = -t_0 \sum_i \left( c_i^\dagger c_{i+1} + \mathrm{h.c.} \right) - g\sqrt{2\omega_0} \sum_i x_i c_i^\dagger c_i .
\end{equation}
Here, the ion position operators $x_i = (1/\sqrt{2\omega_0}) (a_i^\dagger + a_i) $ are replaced by the coordinates. 
%The ion dynamics depends on the initial displacements and velocities. 
If we freeze the ions at their $t=0$ positions, we obtain the Anderson model with diagonal disorder 
\begin{equation}\label{eq: H_AM}
 H^{\mathrm{el}}(0) = -t_0 \sum_i \left( c_i^\dagger c_{i+1} + \mathrm{h.c.} \right) + \sum_i \varepsilon_i c_i^\dagger c_i ,
\end{equation}
where $\varepsilon_i = -g \sqrt{2\omega_0} x_i$. The disorder distribution is Gaussian with the variance given by Eq.~(\ref{eq:variance}), $\sigma_{\varepsilon}^2 = g^2  \mathrm{coth}  \left( \frac{\beta\omega_0}{2} \right) $. The corresponding displacement variance is $\langle x_i^2(0) \rangle = \frac{1}{2\omega_0} \coth \left( {\beta \omega_0} / {2}\right)$. To calculate the current-current correlation function given by Eq.~(\ref{eq_SM_Cjj}), we need the time evolution of the eigenstates $|n\rangle \equiv |\psi_n(0)\rangle$, $H^{\mathrm{el}}(0) |\psi_n(0)\rangle = E_n |\psi_n(0)\rangle$, according to Hamiltonian (\ref{eq: H_el})
\begin{equation}\label{eq:time_evolution}
 i\frac{\partial}{\partial t} |\psi_n(t)\rangle = H^{\mathrm{el}}(t) |\psi_n(t)\rangle .
\end{equation}

The electron impact to the ion dynamics can be either neglected or included through the back-action term in the mean-field Ehrenfest approach. The first approach, called classical path approximation (CPA) {\red{\cite{SMRuneson_2024,SMWang_2011}}}, is slightly simpler to implement since the ion dynamics is completely determined by the initial ion displacements $x_i(0)$ and velocities $\dot x_i(0)$
\begin{equation}\label{eq:CPAions}
 x_i(t) = x_i(0)\cos(\omega_0 t) + (\dot x_i(0)/\omega_0) \sin(\omega_0 t) .
\end{equation}
Analogously to $x_i(0)$, the initial velocities are taken from the Gaussian distribution with the variance $\langle \dot x_i^2(0) \rangle = \frac{\omega_0}{2} \coth \left( {\beta \omega_0} / {2}\right)$.
Then, Eq.~(\ref{eq:time_evolution}) can be solved by the $4^{\mathrm{th}}$ order Runge-Kutta method. We checked that it was sufficient to take the time step $\Delta t=0.05$. To decrease the finite-size effects one can modulate the ion frequency by a randomly chosen $d\omega_{0i}$, see  Sec.~\ref{Sec:SM2B}.

%We also checked that the results do not depend on the size of the lattice, which needs to be much longer than in the fully quantum solution. In our case, the size of the lattice typically had few hundred of lattice sites and we averaged over thousands of initial configurations.

In the Ehrenfest approach, the ion dynamics is modified by the back-action term, which conserves the total energy of the system. (The electron kinetic energy, however, changes with time in both approaches.) The ion equation of motion is given by
\begin{equation}\label{eq:Ehrenfest}
 \ddot x_i(t) = -\omega_0^2 x_i(t) - \frac{\partial}{\partial x_i} \langle \langle \psi_n(t)|H^{\mathrm{el}}|\psi_n(t) \rangle \rangle ,
\end{equation}
where the outer angle bracket denotes the ensemble average, in the spirit of the mean-field Ehrenfest approach \cite{SMTully_2023}. The dynamics does not depend on the ion mass which we set to one. Then, as in Ref.~\cite{SMTroisi_PRL2006} for the Peierls model, by expanding up to the second order and using the Schr\" odinger equation, we get
\begin{eqnarray}\label{eq:Troisi1}
 |\psi_n(t+\Delta t)\rangle &=& |\psi_n(t)\rangle - iH^{\mathrm{el}}|\psi_n(t)\rangle \Delta t \nonumber \\ 
 &-& \! \! \frac{1}{2}i \left[ H^{\mathrm{el}}(t) \dot \psi_n(t) + \dot H^{\mathrm{el}}(t)\psi_n(t)\right] \! \! \Delta t^2 .
\end{eqnarray}
For the Holstein Hamiltonian
\begin{equation}
  \dot H^{\mathrm{el}} = -g\sqrt{2\omega_0} \sum_i \dot x_i c_i^\dagger c_i .
 \end{equation}
The ion positions are updated using the $2^{\mathrm{nd}}$ order Verlet algorithm
\begin{eqnarray}
 x_i(t+\Delta t) = 2x_i(t)-x_i(t-\Delta t) + \ddot x_i(t) \Delta t^2 , \\
 \dot x_i(t+\Delta t) = \frac{1}{2\Delta t} \left[ x_i(t+\Delta t) - x_i(t-\Delta t) \right] .
\end{eqnarray}
The same algorithm can be used for a numerical solution in CPA, where the only difference would be the absence of back-action term
\begin{equation}
 - \frac{\partial}{\partial x_i} \langle \langle \psi_n(t)|H^{\mathrm{el}}|\psi_n(t) \rangle \rangle = \langle \langle \psi_n(t)| g\sqrt{2\omega_0} c_i^\dagger c_i | \psi_n(t) \rangle \rangle .
\end{equation}
In the $2^{\mathrm{nd}}$ order Verlet solution a smaller $\Delta t$ step is needed than in the Runge-Kutta. We checked that setting $\Delta t =0.02$ was enough.

\vspace*{0.5cm}

\section{Quantum-classical dynamics: Selected additional results} \label{SM:glava_druga}

\subsection{CPA vs.~Ehrenfest} \label{Sec:SM2A}

\begin{figure}[h]
\includegraphics[width=\columnwidth]{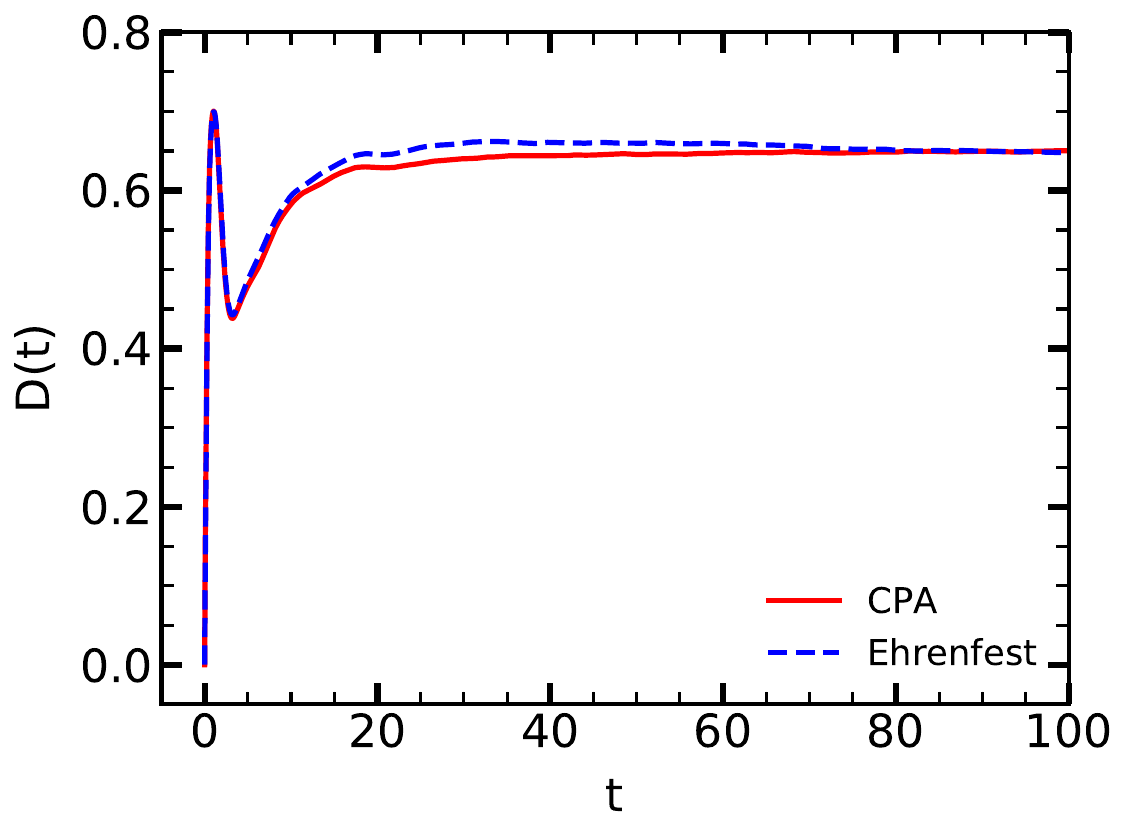}
\caption{Time-dependent diffusion constant in the CPA and Ehrenfest approach. Here $\omega_0=1/3, \lambda=0.5$, and $T=1$.
}
\label{SM_Fig_Ehr}
\end{figure}

The difference between the CPA and the mean-field Ehrenfest method is small in all parameter regimes that we studied. As an illustration, Fig.~\ref{SM_Fig_Ehr} shows the time-dependent diffusion constant for $\omega_0=1/3, \lambda=0.5$ and $T=1$, for the lattice-size $N=200$. The results are averaged over 3000 initial configurations. %, as in Fig.~1 of main text.

\subsection{Finite-size effects} \label{Sec:SM2B}

In the QC approach one needs a longer chain to eliminate the finite-size effects in comparison to a fully quantum solution, where the finite-size effects are diminished by the inelastic electron-phonon scattering. The chain-length in the QC solution needs to be much longer than the localization length (for frozen vibrations). However, for stronger disorder, we also need a rather long chain, to properly sample the tails of the disorder distribution. Typically, we used up to few hundreds of lattice sites and we averaged over thousands of initial configurations.

\begin{figure}[t]
\includegraphics[width=\columnwidth]{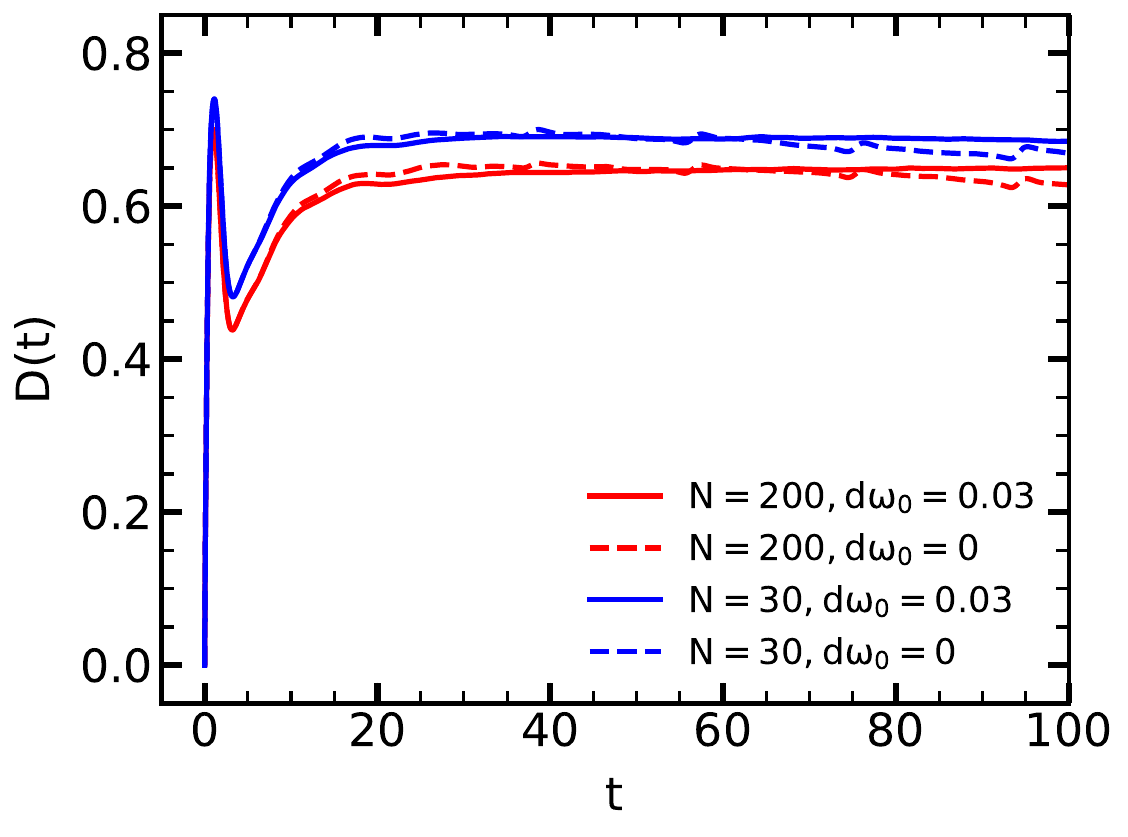}
\caption{Time-dependent diffusion constant for the lattice-
\begin{minipage}[t]{0.49\textwidth}
\justifying $\!\!\!\!\!\!$% Requires \usepackage{ragged2e}
sizes $N=200$ and 30. The solid lines correspond to the case when the ion frequencies are modulated by a random frequency taken from the Gaussian distribution with a standard deviation $d\omega_0=0.03$. Here $\omega_0=1/3, \lambda=0.5$, and $T=1$.
        \end{minipage}
    }

\label{SM_Fig_finite_size}
\end{figure}

We also note that the QC correlation function and time-dependent diffusion constant feature small bumps at the times which are multiple of the period $2\pi/\omega_0$. To eliminate this artifact, we modulate a frequency of the ion $i$ by a small random frequency $d\omega_{0i}$, taken from a Gaussian distribution of standard deviation $d\omega_0$. Such a modification leads to a better formed plateau in $D(t)$ for times $t\gtrsim 2\pi/\omega_0$, as illustrated in SM Fig.~\ref{SM_Fig_finite_size}. We always used a modulation $d\omega_0 \approx \omega_0/10$.

\subsection{Correlation functions on imaginary axis}

Here we show that the difference between the imaginary-axis correlation functions for the Holstein model and the corresponding Anderson model is very small. We explain how such a minuscule difference can still correspond to a significant difference in dc mobilities. 

\begin{figure}[t]
\includegraphics[width=\columnwidth]{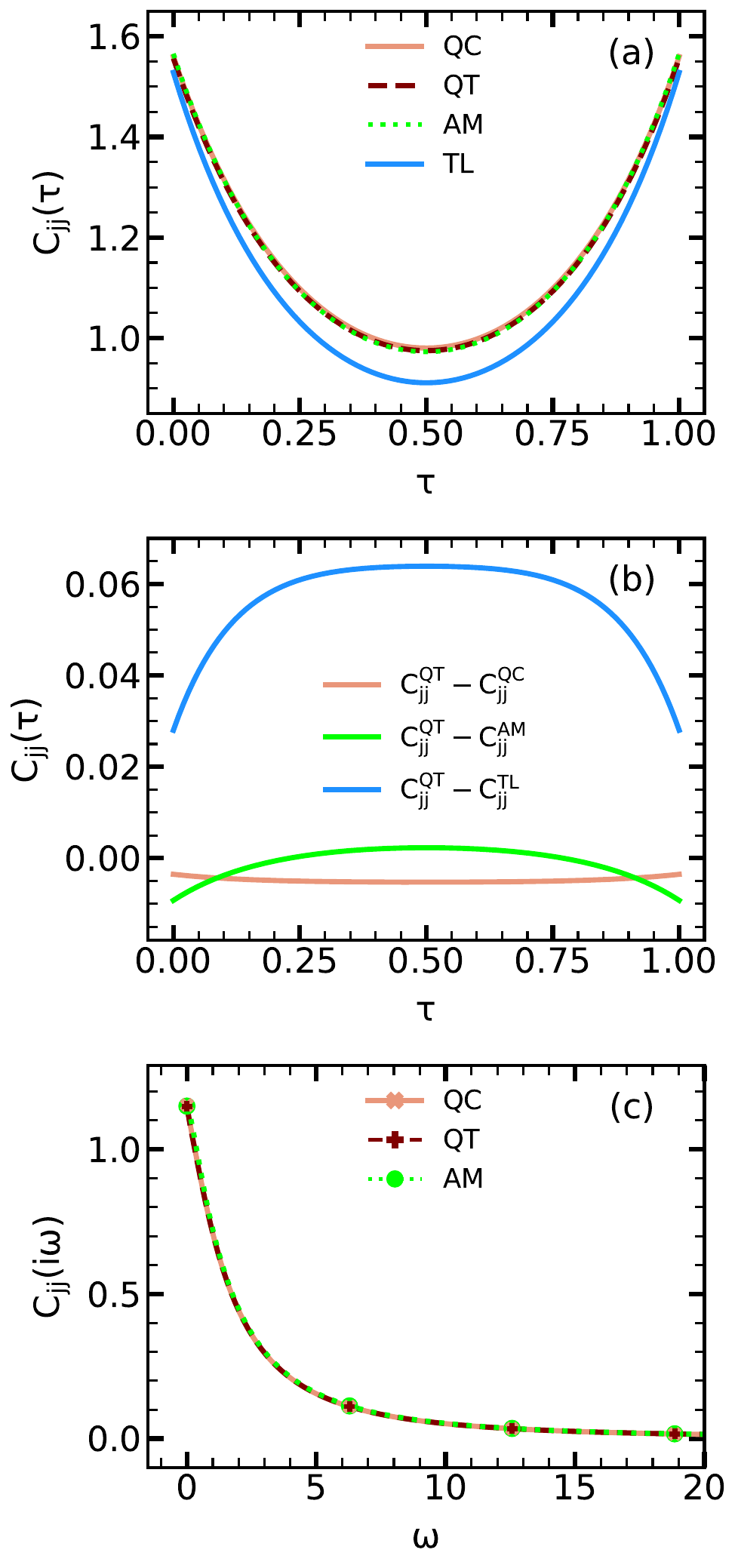}
\caption{(a) Current-current correlation functions in imagi-
\begin{minipage}[t]{0.49\textwidth}
\justifying $\!\!\!\!\!\!$% Requires \usepackage{ragged2e}
nary time, (b) their difference, and (c) the corresponding correlation functions in imaginary frequency. Here $\omega_0=1/3, \lambda=0.5$, and $T=1$.
        \end{minipage}
    }

\label{SM_fig_continuation}
\end{figure}

The imaginary-time correlation function can be obtained from the frequency-dependent mobility as 
\begin{equation}
 C_{jj}(\tau) = \frac{1}{\pi}\int_{-\infty}^{\infty} d\omega \frac{e^{-\tau \omega}}{1-e^{-\beta \omega}} \omega \mu(\omega) .
\end{equation}
We note that $C_{jj}(\tau)$ and $C_{jj}(i\omega)$ are purely real and $\mu(-\omega) = \mu(\omega)$.
In Fig.~\ref{SM_fig_continuation}(a) we show $C_{jj}(\tau)$ for QC, QT, AM and TL solutions for parameters as in Fig.~1 of main text. The QC, QT and AM curves almost overlap. Their difference is smaller than 0.01, as shown in panel (b). The difference with the TL curve is larger due to the difference in $\mu(\omega)$ for  $\omega \gtrsim \omega_0$. 

The current-current correlation function in imaginary frequency can be also easily obtained using the expression
\begin{equation}
 C_{jj}(i\omega) = \frac{1}{\pi}\int_{-\infty}^{\infty} d\nu \frac{\nu^2}{\nu^2 + \omega^2} \mu(\nu) .
\end{equation}
The corresponding QC, QT and AM curves are shown in Fig.~\ref{SM_fig_continuation}(c). The symbols are shown at the Matsubara frequencies $\omega_n = 2n\pi T$. Since the dc mobility is equal to the derivative
\begin{equation}
 \mu_{\mathrm{dc}} = \left. \frac{\partial {\mathrm {Im}} C_{jj}(\omega)}{\partial \omega} \right|_{\omega = 0^+} = - \left. \frac{\partial {\mathrm{Re}} C_{jj}(i\omega)}{\partial \omega} \right|_{\omega=0^+} ,
\end{equation}
it explains how such a tiny difference in $ C_{jj}(i\omega_n)$ at the Matsubara frequencies can still correspond to a subtantial difference in dc mobilities.

\subsection{Electron kinetic energy}

\begin{figure}[t]
\includegraphics[width=\columnwidth]{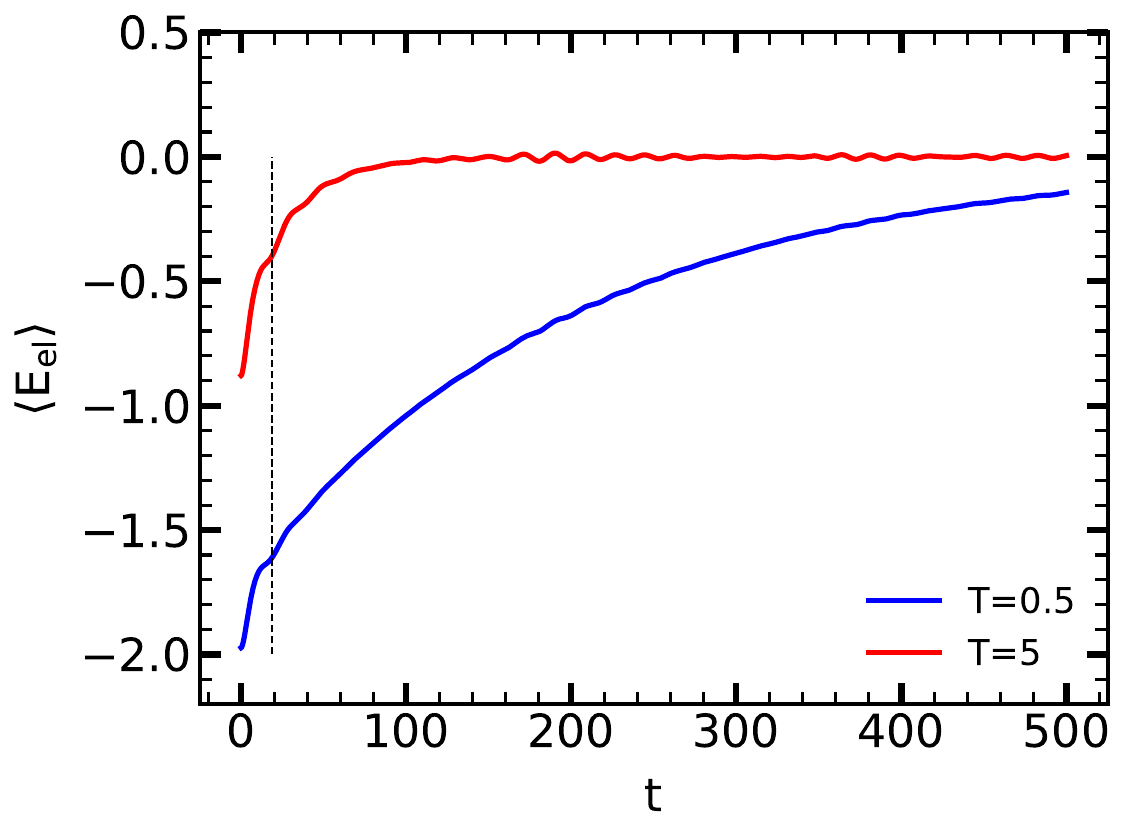}
\caption{Ensemble averaged kinetic energy for $T=0.5$ and 
\begin{minipage}[t]{0.49\textwidth}
\justifying $\!\!\!\!\!\!$% Requires \usepackage{ragged2e}
$T=5$. Here, $\lambda=1/8$ and $\omega_0=1/3$.
        \end{minipage}
    }

\label{SM_Fig_E_el}
\end{figure}

In the QC approach the ensemble-averaged kinetic energy, 
\begin{equation}
 \langle E_{\mathrm{el}} (t) \rangle = \frac{1}{Z} \sum_n \langle \psi_n (t) | H^{\mathrm{el}}(t) |\psi_n(t) \rangle ,
\end{equation}
increases with time, which is a well-known artifact of this method \cite{SMFratini2016}. Still, from the examples where we made a comparison with the quantum solution of the Holstein model, we see that the impact of this artifact on the diffusion $D(t)$ is not drastic. In Fig.~\ref{SM_Fig_E_el} we show $\langle E_{\mathrm{el}} (t) \rangle $ for weak coupling $\lambda=1/8$ and $\omega_0=1/3$. At low temperature $T=0.5$ the dynamical disorder is weak, the approach to the long-time limit $\langle E_{\mathrm{el}} (t) \rangle = 0$ is slow, and $\langle E_{\mathrm{el}} (t) \rangle$ did not significantly change by the time $2\pi/\omega_0$ when the plateau in $D(t)$ is reached. In this case $D(t)$ quantitatively agrees very well with the quantum (DMFT) solution (see main text). At $T=5$ the dynamical disorder is strong and $\langle E_{\mathrm{el}} (t) \rangle$ approaches much faster to the long time limit. Yet, even in this case the QC solution is close to the quantum (HEOM) solution. We note that $\langle E_{\mathrm{el}} (0) \rangle$ is larger for $T=5$ than for $T=0.5$ since in the former case also the states close to the center of the band are partially occupied.

\subsection{Strong electron-phonon coupling} \label{Sec:SM2C}

\begin{figure}[b]
\includegraphics[width=\columnwidth]{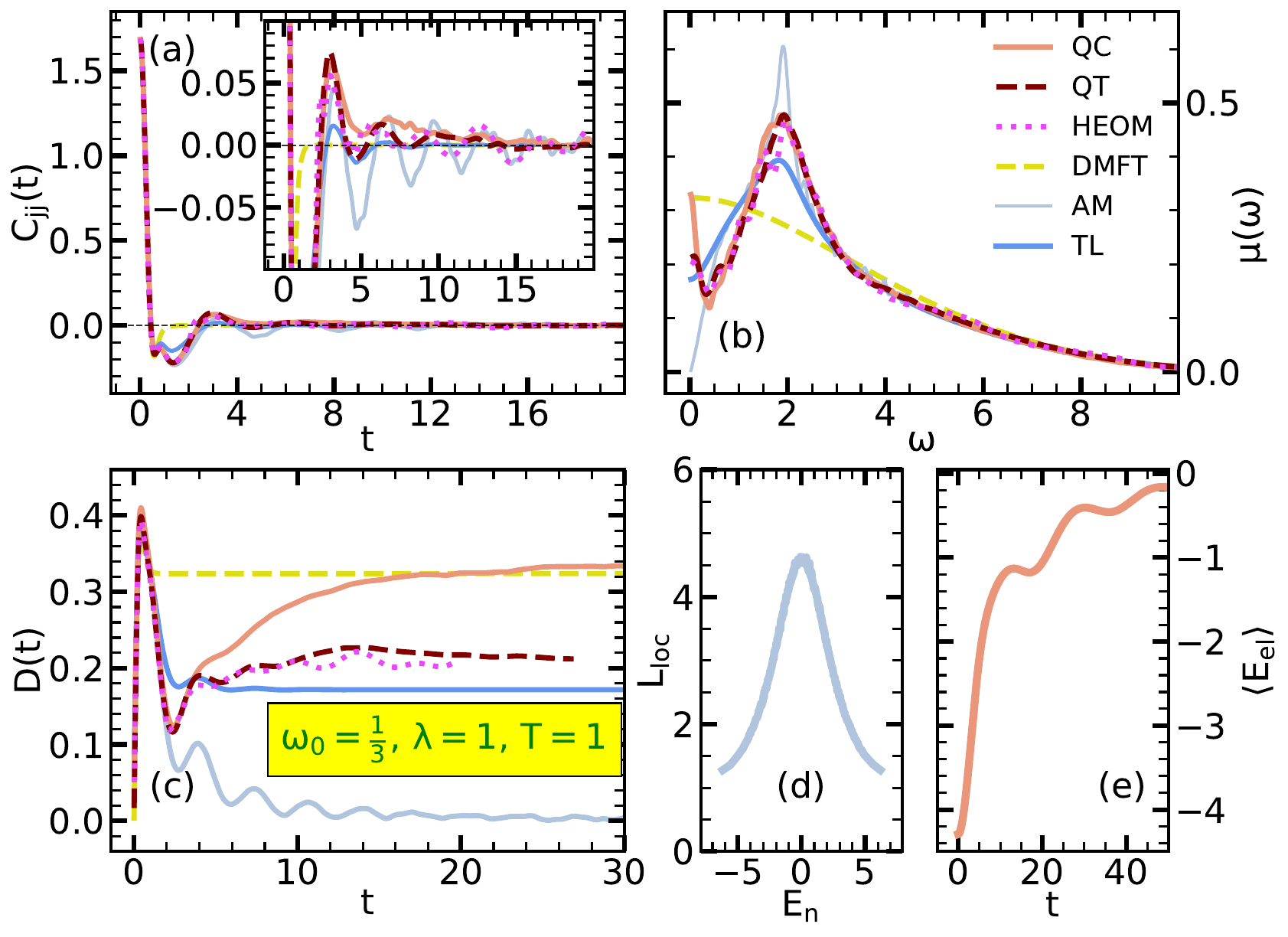}
\caption{Comparison between different methods of the 
\begin{minipage}[t]{0.49\textwidth}
\justifying $\!\!\!\!\!\!$% Requires \usepackage{ragged2e}
(a) current-current correlation function, (b) frequency-dependent mobility, (c) time-dependent diffusion constant, (d) localizaton length and (e) ensemble averaged electron kinetic energy for strong electron-phonon coupling. Here $\omega_0=1/3, \lambda=1$, and $T=1$.
        \end{minipage}
    }
\label{SM_Fig_Rezim_37}
\end{figure}

The results for
%the disorder strength that we refer to as a strong intermediate, with
$\omega_0=1/3, \lambda=1$ and $T=1$, are shown in SM Fig.~\ref{SM_Fig_Rezim_37}. Here we used the chain of the length $N=300$ and 3200 initial condition realizations. As compared to Fig.~1 of main text, here we observe a quantitative difference in $D(t)$ at longer times. This could be a consequence of a substantial change of the electron kinetic energy on the QC classical solution by the time that the plateau is reached in $D(t)$, as a consequence of stronger electron-phonon coupling.

\end{document}